\documentclass{ws-procs961x669}            
\usepackage{hhline}
\begin{document}
\title{A cryogenic and superconducting inertial sensor for the Lunar Gravitational--Wave Antenna, the Einstein Telescope and Selene-physics}

\author{F. Badaracco$^*$}
\author{J.V. van Heijningen}
\author{E.C. Ferreira} 

\address{Centre for Cosmology, Particle Physics and Phenomenology, Université catholique de Louvain,\\
Louvain-La-Neuve, B-1348, Belgium\\
$^*$E-mail: francesca.badaracco@uclouvain.be\\
http://www.uclouvain.be/}

\author{A. Perali}
\address{School of Pharmacy, Physics Unit, SuperNano Laboratory, University of Camerino,\\ 62032 - Camerino (MC), Italy\\
http://www.supermaterials.org}
\address{INAF - Sezione di Camerino, Via Gentile III da Varano, 27, 62032 - Camerino (MC) - Italy}

\begin{abstract}
The Lunar Gravitational--Wave Antenna is a proposed low--frequency gravitational--wave detector on the Moon surface. It will be composed of an array of high-end cryogenic superconducting inertial sensors (CSISs).
A cryogenic environment will be used in combination with superconducting materials to open up pathways to low--loss actuators and sensor mechanics. CSIS revolutionizes the (cryogenic) inertial sensor field with a modelled displacement sensitivity at 0.5\;Hz of 3 orders of magnitude better than the current state--of--the--art. It will allow the Lunar Gravitational--Wave Antenna to be sensitive below 1 Hz, down to 1\;mHz and it will also be employed in the forthcoming Einstein Telescope---a third-generation gravitational--wave detector which will make use of cryogenic technologies and that will have an enhanced sensitivity below 10\;Hz. Moreover, CSIS seismic data could also be employed to obtain new insights about the Moon interior and what we can call the Selene-physics.
\end{abstract}

\keywords{Inertial sensor, Superconducting, Cryogenic, Seismic sensor, Lunar science, Lunar gravitational-wave antenna, Einstein Telescope}

\section{Introduction}
The development of the Cryogenic Superconducting Inertial Sensor (CSIS) originates in the Einstein Telescope (ET) framework. ET will need a highly precise sensor to monitoring motion effects caused by the low--vibration cooling applied to its penultimate suspension stage, which indeed will operate at cryogenic temperatures and in vacuum \cite{ESFRI_long_2020}. Because of its extreme sensitivity and capability of working at cryogenic temperatures, CSIS will also be deployed on the Moon, exploiting it as a detector to reveal gravitational waves (GWs). The Lunar Gravitational--Wave Antenna (LGWA) \cite{Harms_2021} aims to exploit Moon's response to passing GWs and its resulting surface motion. It will employ an array of 4 CSISs which can be regarded as the readout of a detector constituted by the Moon. The seismic data that LGWA would record will be used to study the Moon interior, its seismicity and its formation. This field of study is addressed as Selene-physics (the geophysics of the Moon).

\begin{figure}
\includegraphics[width=1\textwidth]{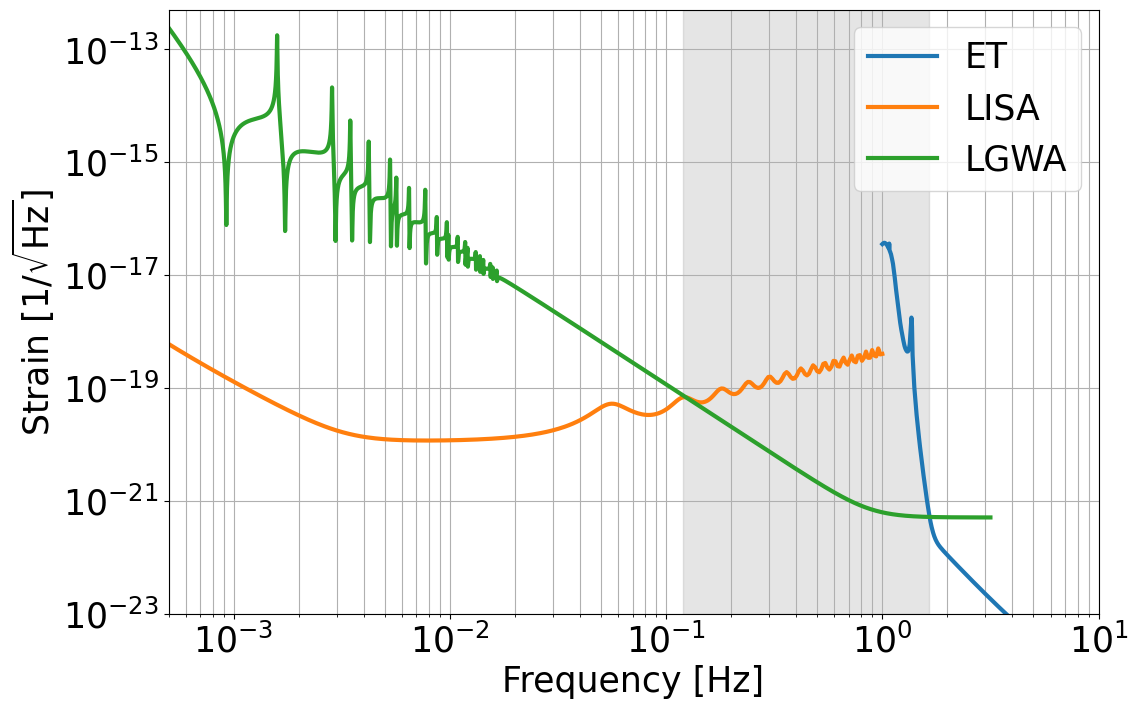}
\caption{Comparison in strain sensitivity between three future GW detectors. The notch--peak structures in the LGWA curve around 1--10\;mHz result from the response of the Moon to passing GWs. Combining the performance of CSIS with the expected surface motion due to the lunar response yields the LGWA strain sensitivity. The grey rectangle highlights the gap that LGWA would bridge. See Ref. \cite{Harms_2021} for more info about the LGWA sensitivity curve.}
\label{fig:LGWA_curve}
\end{figure}

LGWA will bridge the sensitivity gap between ET and the Laser Interferometer Space Antenna (LISA) \cite{Danzmann1996} (see Fig. \ref{fig:LGWA_curve}) between 0.12 and 1.5\;Hz. 

Improving detector sensitivity in the low--frequency band (below 10\;Hz) entails a huge technical effort: Earth--bound GW detectors, like ET, are limited by the Newtonian noise \cite{Harms2019}, while LGWA poses some difficulties deriving from deploying, assembling and powering the necessary instruments on the Moon. \\
The three detectors of which the sensitivity curves are shown in Fig.~\ref{fig:LGWA_curve} are in development to extend our ability to detect GWs at lower frequencies. Signals of binary systems of (super--)massive black holes are found at these frequencies. Additionally, it will allow us to observe binary neutron stars already hours before the final merging \cite{Chan2018,Grimm2020}. Release of early warnings for the Electromagnetic (EM) follow--up will then be possible. The longer the observational time, the better the parameter estimation will be, also allowing to perform the sky--localization solely with the aid of LGWA (by exploiting the rotation of the Moon around the Earth). Polarization measurements and general relativity tests will also be possible \cite{Harms_2021}. Finally, it is impossible to know, but exciting to imagine what kinds of unknown unknowns can be detected once we open up that low--frequency band.

\section{A technical design for CSIS}\label{sec:CSIS}
\begin{figure}
\includegraphics[width=1\textwidth]{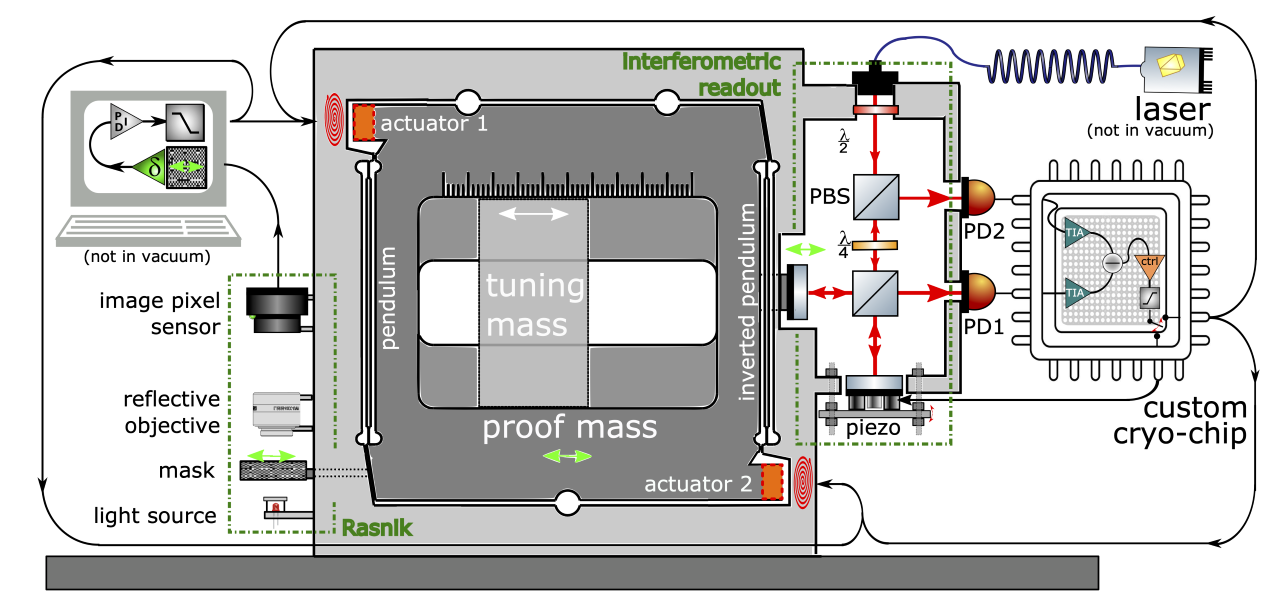}
\caption{Design of the CSIS sensor. We can identify four main parts: two readouts (highlighted by green dotted rectangles), spiral actuators (orange and red) and the proof mass (at the center). In Sec. \ref{sec:CSIS} more details about the actuators and the proof mass are provided, while the two readouts are described in Secs. \ref{sec:itf_readout} and \ref{sec:rasnik_readout}.}
\label{fig:CSIS}
\end{figure}
The design of CSIS is presented in Fig.~\ref{fig:CSIS}. We can identify its four main parts: the Rasnik readout, the interferometric readouts, the two actuators on either side of the proof mass, and the niobium monolithic sensor mechanics. The latter will be the core of our sensor: a 1\;kg mass suspended in a Watt's linkage configuration~\cite{Liu1997} (a combination of a pendulum and an inverted one) fabricated by Electrical Discharge Machining (EDM). The monolithic design allows to avoid having separate mechanical parts which would cause thermal dissipation~\cite{Bertolini2006}. 
\begin{figure}
\includegraphics[width=1\textwidth]{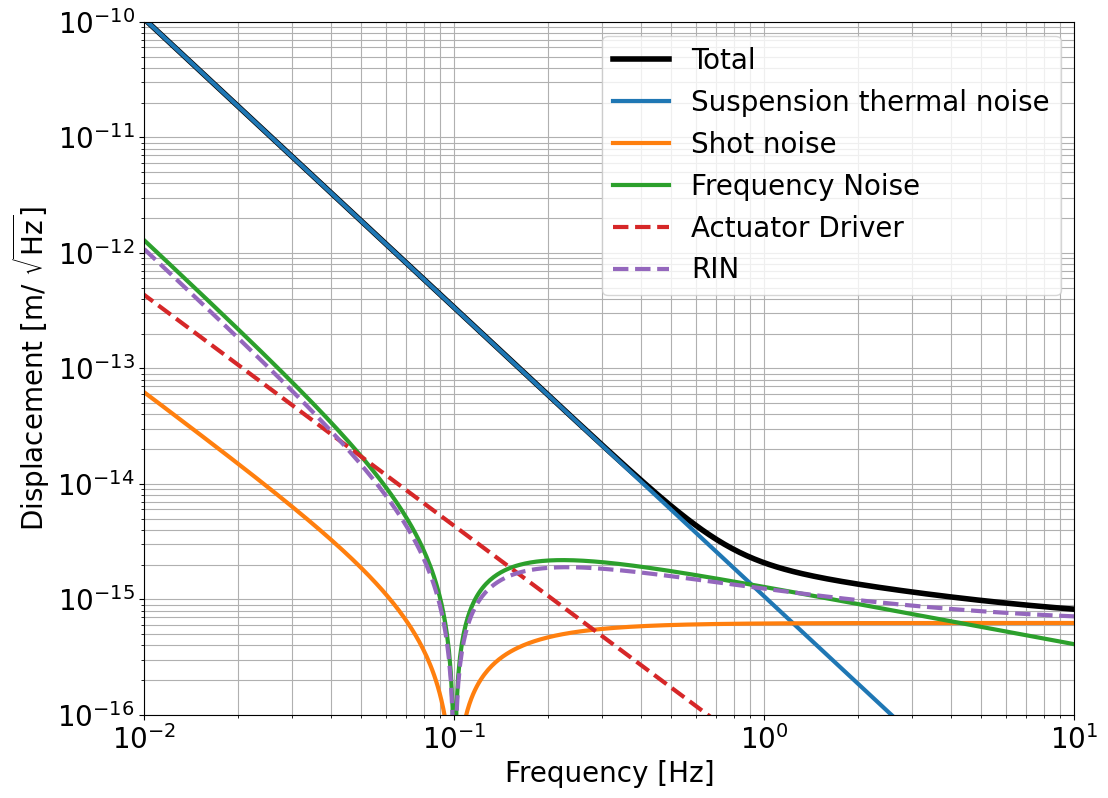}
\caption{Sensitivity and relative noises of niobium CSIS. The thermal noise was calculated considering a Q--factor of 10$^\text{4}$ and a temperature of 5\;K. The other noises where calculated based on Ref.\cite{Heijningen2020}, where RIN is referring to the Relative Intensity Noise of the laser source of the interferometric readout and the actuator driver is calculated following Eq. \ref{eq:AD}.}
\label{fig:noises}
\end{figure}
Moreover, in Fig.~\ref{fig:CSIS} we see a tuning mass that will serve to adjust the resonance frequency of the sensor. 
\begin{figure}
\includegraphics[width=1\textwidth]{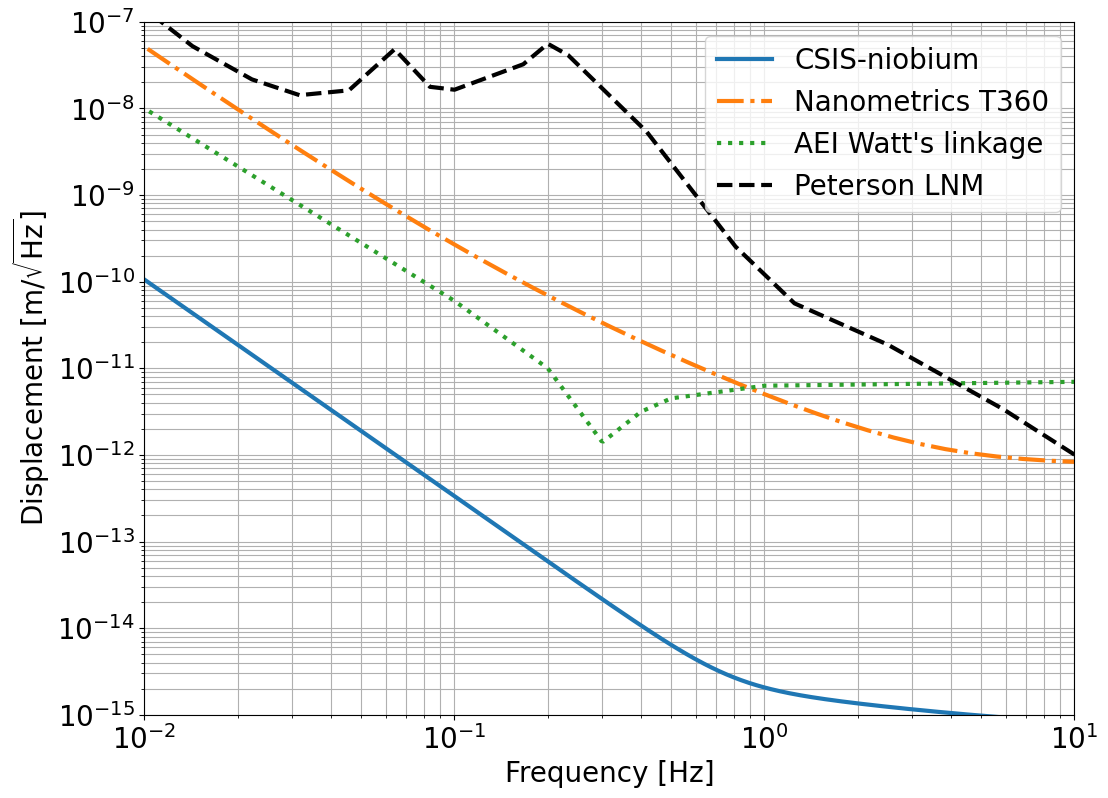}
\caption{Sensitivity of CSIS made of niobium  compared to that of the most sensitive seismic sensors available. The Peterson’s Low Noise Model (LNM) is depicted for comparison. }
\label{fig:comparison}
\end{figure}

At cryogenic temperatures we can dramatically reduce the thermal noise, especially by employing materials that are superconductive (T$_\text{c}$ = 9.2\;K). At $\sim 5\;$K the mechanical Q--factor of a niobium CSIS will be of the order of 10$^{\text{4}}$ (see Ref.\cite{Heijningen2020}), which is a significant improvement with respect to the room temperature version. Concerning LGWA, permanent shadow zones (of the order of several km) on the Moon are a convenient way to provide low and stable temperatures ($<$40\;K )\cite{Paige2010}. This will make less challenging reaching the temperatures where the niobium behaves as a superconductor.

The actuators of Fig. \ref{fig:CSIS} will constitute of a thin layer of niobium deposited on a ceramic substrate and then attached to the sensor frame. By applying a current to the spiral coil, a magnetic field will be created. Thanks to the Meissner effect, it will be expelled from the niobium of the proof mass, thus exerting a force on it.\\ To calculate CSIS sensitivity of Fig. \ref{fig:noises}, we need to take into account also the noise produced by the actuator driver:
\begin{equation}\label{eq:AD}
n_{\text{AD}} = \frac{\beta \text{V}_{\text{DAC}}}{\text{R}_{\text{s}} \text {m} \omega^2},
\end{equation}
where m is the mass, $\text{R}_{\text{s}}$ is the sampling resistance (which determines the current into the actuator), $\text{V}_{\text{DAC}}$ is the DAC voltage noise and $\beta$ is the coil response value based on the latest simulations. 

Another important advantage of employing a superconducting material is that it will avoid the thermal dissipation due to the eddy currents. This was a significant problem for the room temperature sensor version, where voice coils were employed to actuate the proof mass. In this way, the mechanical Q--factor was found to be highly limited by viscous damping associated with eddy currents induced on the moving metal surfaces by the voice coil stray field. The two readouts in Fig. \ref{fig:CSIS} will be instead described in the next two sections. 

Thanks to its monolithic design, the employment of a material with a high Q-factor and thanks to two different readouts, CSIS will be able to reach sensitivities of 2\;fm/$\sqrt{\text{Hz}}$ at 1 Hz (see Fig. \ref{fig:noises}). This is a factor 3000 better than the most sensitive seismic sensors (see Fig. \ref{fig:comparison}, where the Peterson’s Low Noise Model (LNM) \cite{Pet1993} is depicted for comparison).

\section{Interferometric readout}\label{sec:itf_readout}
\begin{figure}
\centering
\includegraphics[width=0.5\textwidth]{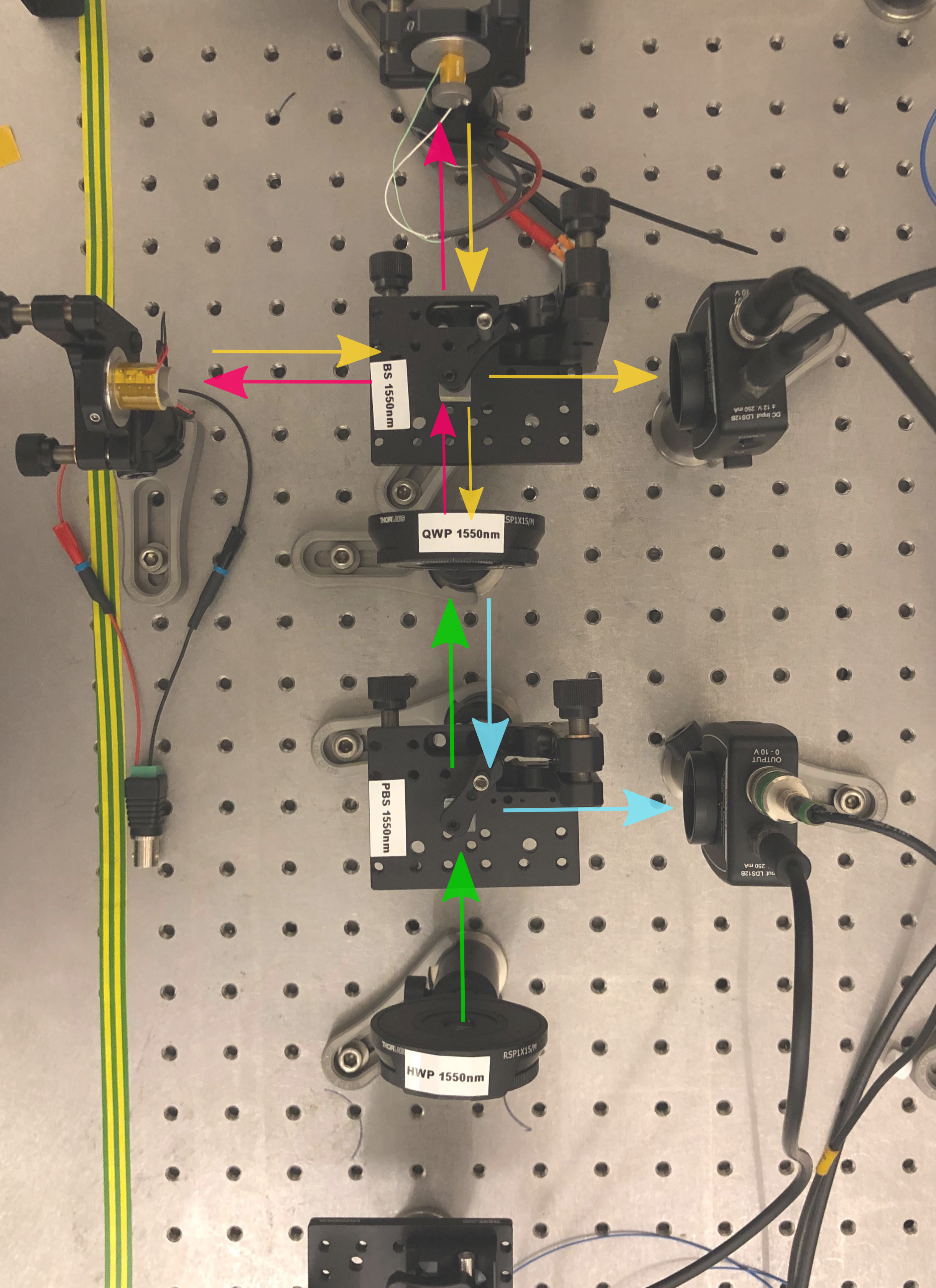}
\caption{Readout interferometer set up (laser on the bottom and photodiodes on the right). The arrows represent the light polarization: the green arrow right after the half wave plate (AHWP05M-1600-Thorlabs) is the linearly polarized light along the transmission axis of the polarizing beamsplitter (PBS104-Thorlabs), which, after passing through a quarter wave plate (AQWP05M-1600-Thorlabs), becomes circularly polarized (pink arrow). Reflection on the mirrors then reverses the handedness (yellow arrow). The second passage through the quarter wave plate with reversed handedness produces a linear polarization rotated by 90° degrees, which allows the light to be reflected by the polarizing beamsplitter instead of being transmitted. }
\label{fig:ITF}
\end{figure}

The interferometric readout that will be implemented in CSIS is based on that of Ref. \cite{Gray1999} and it can be seen on the right side of Fig. \ref{fig:CSIS} and in Fig. \ref{fig:ITF}. It can be described as a simple Michelson interferometer with one more beamsplitter that allows to read the light coming back towards the laser. This second beamsplitter is needed to remove the common noise (such as intensity fluctuations in the laser power) by taking the differential signal between the two outputs. In this way, we can reach a shot noise limited sensitivity. The subtraction will be performed by a chip designed to work at cryogenic temperatures using CMOS technology. 

\begin{figure}[htbp]
\includegraphics[width=1\textwidth]{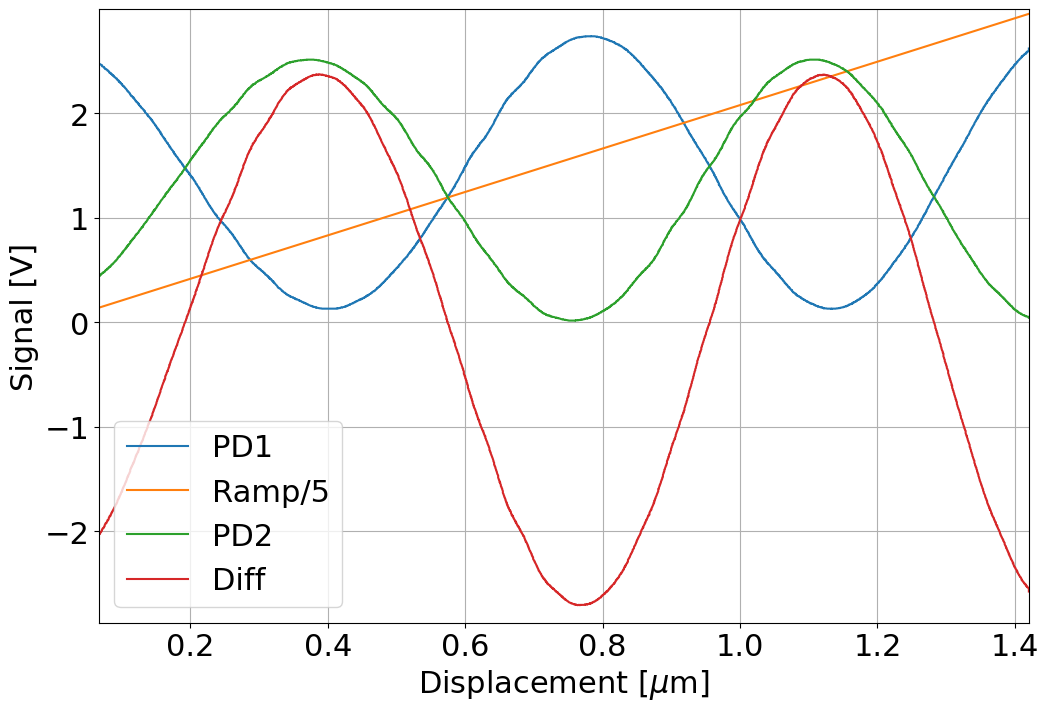}
\caption{Interference fringes obtained with a 1538\;nm laser and moving one of the mirrors by injecting a ramp signal in its piezo actuator (PK44M3B8P2-Thorlabs). Using the known wavelength we find a piezo conversion factor of 96.1\;nm\,V$^{-1}$. The minima and maxima are not exactly separated by the same distance probably due to non-linearities of the piezo actuator. The working points of the control loop are at 0\;V of the differential signal. }
\label{fig:fringes}
\end{figure}

The differential signal (see Fig. \ref{fig:fringes}) will be fed through a control filter to the actuators which will dynamically lock the proof mass. This results in highly reduced relative motion between the proof mass and its frame and increased readout dynamic range. Moreover, keeping the differential signal as close as possible to a fixed working point ensures optimal subtraction of the common noises. Indeed, in the room temperature version, it was found that the the desired shot-noise-limit performance was not yet achieved~\cite{Joris_PhD}. The reason was that there was still some residual motion of the proof mass, thus resulting in a sub-optimal subtraction of the two photodiodes signals. To solve this, we can use a second control loop to reduce as much as possible the residual motion of the proof mass: the Rasnik readout. 

Polarizing optics divert all the available power onto the photodiodes, avoiding dumping part of it. This is a important feature when working at cryogenic temperatures and it also means less shot noise for given injected power. Indeed, in Tab.~\ref{tab:pd} we can compare the powers received by the photodiodes and the overall shot noise when polarizing and non-polarizing optics are used. The total shot noise expected in the interferometer output depends on the amount of power falling on the photodiodes as $\text{SN}_{\text{i}}\propto 1/\sqrt{P_{\text{i}}}$. The total shot noise is then $\text{SN}_{\text{tot}} = \sqrt{\text{SN}_{\text {PD}_{\text{1}}}^2+\text{SN}_{\text {PD}_{\text{2}}}^2}$.
\begin{table}
\tbl{Comparison of interferometric readout outputs and shot noise without and with polarizing optics for a given input power $P_{\text{0}}$.}
{\begin{tabular}{@{}l l l l @{}}
              ~~&power on PD1              ~~&power on PD2             ~~&~~ $\text{SN}_{\text{tot}}$~~~ \\ 
\hline \hline
~~previous configuration~\cite{Gray1999} ~ & $P_{\text{0}}/4$ ~~& $P_{\text{0}}/8$ ~~&~~ $\propto\sqrt{12/P_{\text{0}}}$~~\\ 

~~with polarizing optics~~ & $P_{\text{0}}/2$ ~~& $P_{\text{0}}/2$ ~~&~~ $\propto\sqrt{4/P_{\text{0}}}$~~\\ 
\hline
\end{tabular}}\label{tab:pd}
\end{table}\\
We thus get a factor $\sqrt{3}$ of reduction in the shot noise when polarizing optics are employed with respect to the non-polarizing optics case. 

\section{Rasnik readout}\label{sec:rasnik_readout}
The control loop driven by the Rasnik readout will have a larger dynamic range, capable of damping the proof mass resonance and thus reducing the proof mass residual motion. A Rasnik uses a back-lit chessboard mask which has its image projected via an objective on a pixel camera. The mask moves with the proof mass to which it is attached. The image that is recorded by the camera is then fed to a processing unit which performs a 2-dimensional Fourier transform. This transform will show a peak because of the many periodically spaced black-white transitions in the chessboard image. This peak shifts according to the motion of the mask and a peak fitting algorithm returns then the Rasnik displacement measurement. The obtained sensitivity is 7 pm/$\sqrt{\mathrm{Hz}}$ and a sub-pm/$\sqrt{\mathrm{Hz}}$ sensitivity is expected given the implementation of certain improvements.~\cite{Graaf2021} The beauty of this system is that its dynamic range is only dependent on the chessboard mask dimensions, which can be designed arbitrarily large.  

\section{Final remarks}
The development of highly sensitive inertial sensors in the GW community opens the path to new GW detector concepts like LGWA, but also to the possibility of studying the Moon in more detail, giving rise to Selene-physics. \\
CSIS is the product of R\&D in the ET community and it is the perfect candidate to be employed as readout in LGWA. It will reach a sensitivity of a few fm/$\sqrt{\text{Hz}}$ from 0.5\;Hz onwards and, therefore, it will become the most sensitive inertial sensor capable of working at cryogenic temperatures in the low--frequency band.

\bibliographystyle{ws-procs961x669}
\bibliography{ws-pro-sample}
\end{document}